\documentclass[preprint,showpacs,preprintnumbers,amsmath,amssymb,aps,prb,floatfix]{revtex4-1}
\usepackage{graphicx}
\usepackage{bm}
\usepackage{color}
\def\SiN{Si$_{3+x}$N$_{4-x}$ }
\begin{document}
\newlength{\picwidth}
\setlength{\picwidth}{\textwidth}

\title{Unusual resistance-voltage dependence of nanojunctions during electromigration in ultra-high vacuum}

\author{D. St\"{o}ffler}
\affiliation{Karlsruhe Institute of Technology, Physikalisches Institut and DFG-Center for Functional Nanostructures, P.O.~Box 6980, D-76049 Karlsruhe, Germany}

\author{M. Marz}
\email{michael.marz@kit.edu}
\affiliation{Karlsruhe Institute of Technology, Physikalisches Institut and DFG-Center for Functional Nanostructures, P.O.~Box 6980, D-76049 Karlsruhe, Germany}

\author{B. Kie\ss ig}
\affiliation{Karlsruhe Institute of Technology, Institut f\"{u}r Festk\"{o}rperphysik, P.O.~Box 3640, D-76021 Karlsruhe, Germany}

\author{T. Tomanic}
\affiliation{Karlsruhe Institute of Technology, Physikalisches Institut and DFG-Center for Functional Nanostructures, P.O.~Box 6980, D-76049 Karlsruhe, Germany}

\author{R. Sch\"{a}fer}
\affiliation{Karlsruhe Institute of Technology, Institut f\"{u}r Festk\"{o}rperphysik, P.O.~Box 3640, D-76021 Karlsruhe, Germany}

\author{H. v.\ L\"{o}hneysen}
\affiliation{Karlsruhe Institute of Technology, Physikalisches Institut and DFG-Center for Functional Nanostructures, P.O.~Box 6980, D-76049 Karlsruhe, Germany}
\affiliation{Karlsruhe Institute of Technology, Institut f\"{u}r Festk\"{o}rperphysik, P.O.~Box 3640, D-76021 Karlsruhe, Germany}

\author{R. Hoffmann-Vogel}
\affiliation{Karlsruhe Institute of Technology, Physikalisches Institut and DFG-Center for Functional Nanostructures, P.O.~Box 6980, D-76049 Karlsruhe, Germany}

\date{\today}

\begin{abstract}
The electrical resistance $R$ of metallic nanocontacts subjected to controlled cyclic electromigration in ultra-high vacuum  has been investigated {\it in-situ} as a function of applied voltage $V$. For sufficiently small contacts, i.e., large resistance, a decrease of $R(V)$ while increasing $V$ is observed. This effect is tentatively attributed to the presence of contacts separated by thin vacuum barriers in parallel to ohmic nanocontacts. Simple model calculations indicate that both thermal activation or tunneling can lead to this unusual behavior. We describe our data by a tunneling model whose key parameter, i.e., the tunneling distance, changes because of thermal expansion due to Joule heating and/or electrostatic strain arising from the applied voltage. Oxygen exposure during electromigration prevents the formation of negative $R(V)$ slopes, and at the same time enhances the probability of uncontrolled melting, while other gases show little effects. In addition, indication for field emission has been observed in some samples. 
\end{abstract}
\maketitle
Electronic transport through metallic nanostructures has been of interest for many years. In break-junction experiments where the metallic contact is elongated and thinned as the leads are mechanically pulled  apart~\cite{Agrait03}, the conductance $G$ does not vary continuously with the dimensions as expected, but changes stepwise. These jumps in the junction conductance reflect the atomistic nature and the electron-waveguide properties of the metal. A promising alternative to mechanically controlled break-junctions is controlled electromigration (EM)~\cite{Strachan05,Esen05}. This technique allows reaching the ballistic regime as well~\cite{Hoffmann08}. In the EM process, Joule heating locally enhances the diffusivity of atoms near a preformed notch. Electrostatic and wind forces cause a directional diffusion with a net atom diffusion either along or opposite to the electric current direction, depending on which force is dominant. Primarily, atoms at grain boundaries are known to move. Consequently, the resulting gap will depend on the microstructure of the wire at the beginning of the EM process~\cite{Stoeffler12}. The structural characterization of the contacts is difficult since scanning electron microscopy (SEM) does not yield images with atomic resolution~\cite{Kaspers09}. Transmission electron microscopy (TEM), on the other hand, is difficult to combine with {\it in-situ} preparation techniques~\cite{Strachan06}. Scanning force microscope (SFM) tips are usually not sufficiently sharp to measure the atomic structure during the formation of nm-sized gap. Therefore, electronic transport measurements are often the only possible means to obtain (indirect) information about the contact region.

In the EM process, the voltage dependence of the resistance $R(V)$ is expected to be constant or have a positive slope due to Joule heating. Here, we report on the observation of negative slopes in $R(V)$ that develop in metallic contacts thinned by electromigration under ultra-high vacuum (UHV). These negative slopes are found in different metals and for different substrates. The unusual effect may be attributed to the presence of closely spaced electrodes in parallel to the remaining metallic contact which may allow electron transport by thermal activation or tunneling. Furthermore, the influence of gas dosage, in particular O$_2$, on the electromigration process is discussed.

The starting point for the electromigration thinning process requires metallic nanobridges. We employed different techniques for sample preparation: electron-beam lithography (EBL) with standard lift-off procedure, or shadow evaporation using either transferable \SiN membranes or on-chip microstructured \SiN membranes as masks. The standard lift-off procedure for EBL samples usually leaves molecular residues on the sample surface~\cite{stoeffler11p1}. This 'dirt' might influence EM thinning either by pinning surface diffusion or by enhancing diffusion by surfactant effects. Shadow-evaporation techniques, on the other hand, allow to prepare UHV-clean samples. The nanobridges had typical dimensions of $150-250\,$nm in width, $20-30\,$nm in thickness, and $500-800\,$nm in length, resulting in an initial resistance of $10-40\,\mathrm{\Omega}$. Details of the sample preparation methods are given in the Supplemental Material including a sketch of the sample layout and a list of the sample details~\cite{supplemental}.

A controlled cyclic EM thinning process in UHV was applied to the samples. For this process we record the current $I$ as a function of applied bias voltage $V$ and obtain $R(V)=I/V$, as described in detail in Ref.~\onlinecite{Hoffmann08}. Usually $R(V)$ displays a positive slope corresponding to a negative curvature in the corresponding $I(V)$ characteristic (see, e.\,g. Fig. 1(d)), as expected for Joule heating of a metallic sample. The regime of moderate voltage is characterized by reversible $R(V)$ curves; experiments with decreasing bias within a cycle show the same $R(V)$ dependence as those with increasing bias. At higher voltage, structural changes of the nanobridge are usually signaled by a rapid and irreversible change of $R(V)$. As the nanobridge is heated, EM is enhanced due to thermal activation of atomic diffusion. In order to prevent overheating that could lead to an uncontrolled burn-through of the nanowire, the voltage is switched off as soon as the resistance change has reached a preset percentage $q$ of the initial value. In general, the resistance increases cycle by cycle---indicating successful thinning by EM---due to the thinning of the contact, if $q$ is set appropriately.
 
\begin{figure*}[htb]
\includegraphics[width=\picwidth]{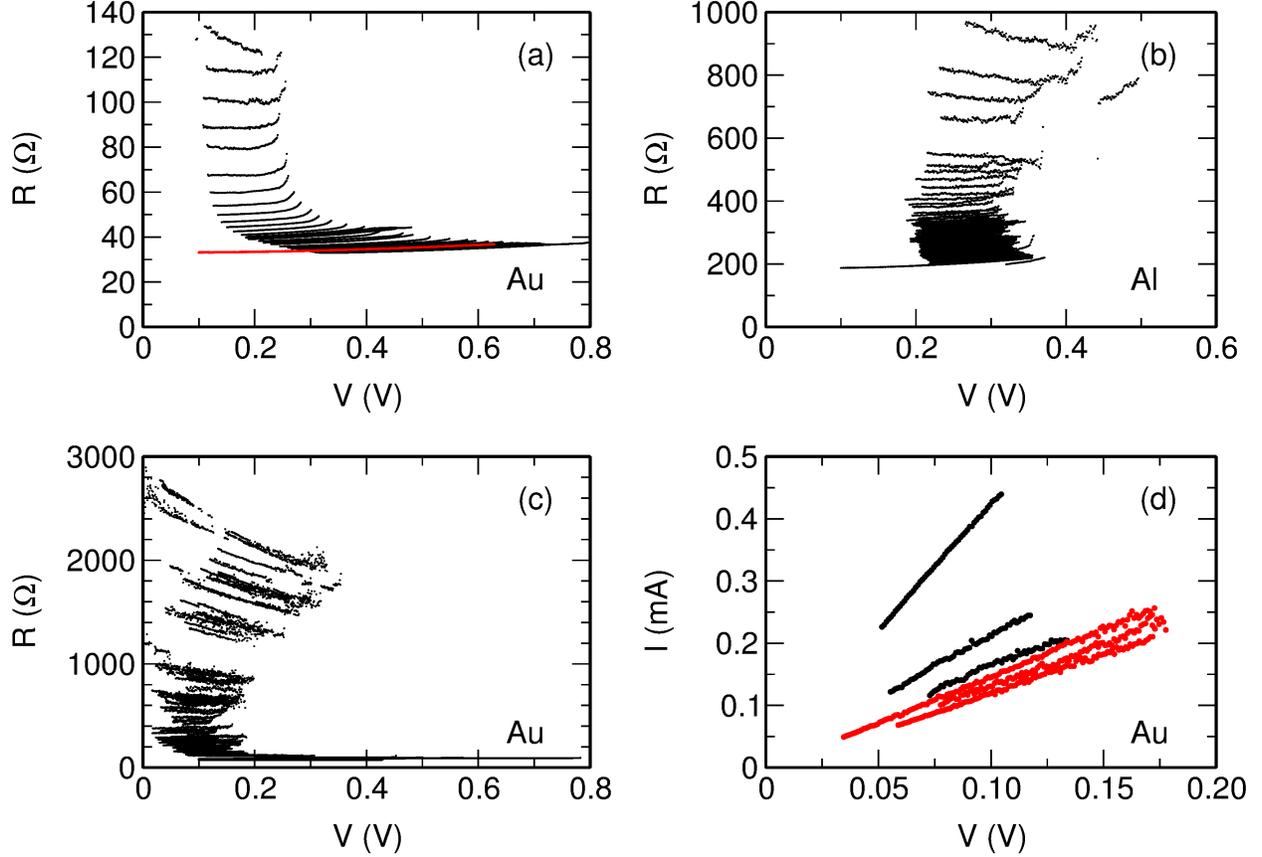}
    \caption{Electromigration thinning of three different samples (selected cycles) in ultra-high vacuum. (a) Shadow-evaporated Au sample, $T= 300\,$K. The first cycle (red) shows a different behavior than the following ones, this is ascribed to annealing of grain boundaries and defects. (b) Lithographically evaporated Al sample; $T= 300\,$K. (c) Mask-evaporated Au sample, EM performed at $T= 4.2\,$K. After several cycles the positive $R(V)$ slope changes to a negative slope. The transition from positive to negative slope occurs either suddenly between subsequent cycles like in (a) or in a continuous fashion, as shown in (b) and (c). (d) Selected current-voltage curves of the sample shown in (c) in the transition region from positive (black) to negative slope (red) in $R(V)$  are depicted. The cycles in black (red) show a slight downward (upward) bending curvature.}
    \label{fig1}
\end{figure*}

The thinning process in UHV usually starts with values of $R(V)$ that are comparable with the ones observed at ambient conditions\cite{Hoffmann08}. After several cycles, in contrast to previous experiments performed in air, the observed positive slope in R(V) changes to a negative one. The crossover from positive to negative slope occurs in some sample in a continuous fashion (see, e.\,g.\ Fig. 1(c)), while in others it occurs rather abruptly from one cycle to the next (see, e.\,g.\ Fig 1(a)). The resistance at which the crossover takes place varies over a range from $150$ to $600\,\mathrm{\Omega}$. The development of a negative slope often prevents the usual cyclic electromigration-thinning process, because the criterion for finishing one EM cycle, i.\,e., an increase of $R$ with respect to its initial value, can not be met at reasonable bias voltage. However, structural changes of the forming constriction within the nanowire are signaled by a rapid upturn of $R(V)$ at the end of a cycle (compare Fig.~\ref{fig1}). We thus modified the halting criterion of the cycle by introducing an additional criterion of a maximally allowed decrease of the resistance during one cycle. Alternatively a differential change of the resistance during consecutive values within a cycle was used to terminate the cycle. This allowed to circumvent the problem partly and enabled controlled electromigration in UHV.

Fig.~\ref{fig1}(a)-(c) shows selected cycles of the EM process for different samples. In each case a crossover of $R(V)$ from a positive to a negative slope was observed. This sign change of the $R(V)$ slope was found in Au, Pt and Al samples, and therefore appears to be independent of the material used. The data show that the occurrence of the negative slope does not depend on the preparation method, on the temperature during the electromigration, the polarity of the electric current, or on the measurement setup. A significant effect of leakage currents through the substrate can be excluded, since similar effects were observed in samples with a thick silicon oxide layer ($\approx 400\,$nm), see e.g., Fig.~\ref{fig1}(a). SEM investigation did not show any significant structural differences, e.g., enhanced crystallite growth, between EM-processed contacts of ambient and UHV-migrated samples.

Scanning force microscopy (SFM) images indicate that due to the EM process, gaps on the order of a few $\mathrm{nm}$ are created in the contacts~\cite{Stoeffler12}. Hence, it seems likely that the structure of the constriction just before complete breaking might be composed of a metallic contact in parallel with a single or a few contacts with a very thin vacuum or oxide barrier, with roughly the same conductance as the remaining metallic contact. Electrons can cross these gaps directly by thermal activation or via tunneling. A sketch of a possible contact configuration is shown in the Supplemental Material (Appendix~C, Fig.~S2)~\cite{supplemental}.  

It has been reported that the EM process can result in clusters of Au containing only $18-22$ atoms, which show Coulomb-blockade effects~\cite{houck05p1}. One therefore might view the contact as being composed of some 'granular' material where the granularity can  origin from insulating barriers or vacuum barriers. The crossover from a positive to a negative slope in resistance could then be explained by thermally activated conduction. However, granular films usually exhibit a typical sheet resistance of several M$\mathrm{\Omega}$ that even increases on annealing~\cite{Barwinski85}. Moreover, our SFM measurements do not show any indication of a granularity in the films~\cite{Stoeffler12}. Therefore, a granular film as cause of the negative slopes in $R(V)$ seems unlikely.

In the following, we will focus on a possible tunneling mechanism as the origin of the negative slopes in $R(V)$. Usually, tunneling contacts have resistances in the range of several tens of k$\mathrm{\Omega}$ to G$\mathrm{\Omega}$. The resistances observed in most of our measurements are typically far below the inverse conductance quantum $G_0^{-1}\approx 13\,$k$\mathrm{\Omega}$,i.e.~the order of magnitude for a single-atom contact. This means that the metallic contact must still be partly connected, or/and that tunneling is effective over a large area.

The proposed model, assuming the presence of tunneling in parallel to a metallic contact, is in agreement with our studies on the morphology changes occurring during EM~\cite{Stoeffler12,stoefflerphd}. We found that the initial grain size ($\approx$ several tens of nm) remains constant during the process and a narrow slit is formed~\cite{Stoeffler12}.
In contrast, recrystallization accompanied by the formation of a large gap where tunneling might be excluded has been observed in recent TEM studies of Au~\cite{strachan08} and Pd/Pt~\cite{kozlova13} contacts. These differences could arise from the distinct substrates and EM materials used, and also from the different observation methods (TEM as compared to SFM). The influence of the substrate material on the gap size was demonstrated by Strachan {\it et al.}~\cite{Strachan06} with TEM. They found a larger gap size for samples prepared on SiN membranes than for samples on Si substrates~\cite{Strachan06}. In general, a direct comparison of TEM studies to our experiment is difficult since for {\it in-situ} TEM studies, samples have to be prepared on very thin substrates. The thickness of the substrate, by virtue of its thermal conductance, is known to have a significant influence on the effective temperature profile along the nanobridge during the EM~\cite{durkan00, kiessig13}. This temperature profile is decisive because different temperatures during the process may result in different structural changes and gap formations. Concerning different materials, we want to point out that void formation has been observed for Pd/Pt alloys~\cite{kozlova13}, which is in contrast to our SFM studies of pure Au samples~\cite{Stoeffler12}.

\begin{figure*}[ht]
\includegraphics[width=\picwidth]{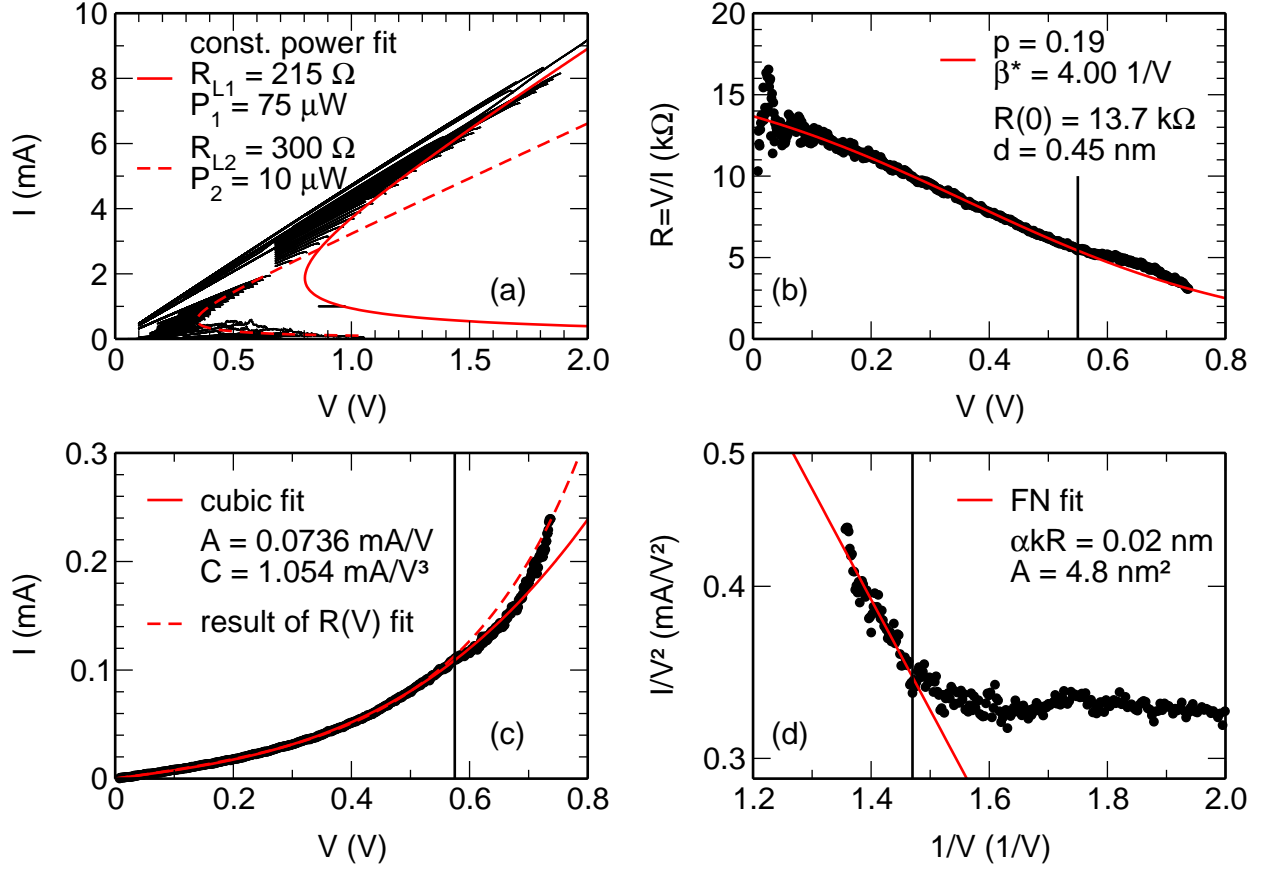}
    \caption{(a) $I$ vs.~$V$ cycles of the full EM on a Au sample (number 4 in the list of samples in the Supplemental Material~\cite{supplemental}). Fit of two lines of constant power (red), given by $V(I)=P/I+R_{\mathrm{L}}\cdot I$, where $P$ is the power where a preset resistance increase during a cycle indicates EM and starts a new cycle, and $R_{\mathrm{L}}$ the lead resistance~\cite{Esen05,Hoffmann08,Stoeffler12}. The fits result in $P_1=75\,\mathrm{\mu W}$ with $R_{\mathrm{L1}}=215\,\mathrm{\Omega}$, and $P_2=10\,\mathrm{\mu W}$ with $R_{\mathrm{L2}}=300\,\mathrm{\Omega}$. (b) Single cycle at high resistance $R(0)\approx 13\,\mathrm{k\Omega}$  exhibiting a negative slope in $R(V)$. The fit (red) was obtained by a modified tunneling model including thermal expansion. For the fit we used the data with $V\lesssim 0.6\,\mathrm{V}$ (vertical line) and also neglected the strong scattered data for low bias, yielding $p=0.19$ and $d=0.45\,$nm (c) Corresponding $I(V)$ measurement with cubic fit (solid red line). For the fit we again used the data with $V\lesssim 0.6\,\mathrm{V}$ (vertical line). The dashed line shows the result of $I(V)$ calculated from the parameters obtained from the $R(V)$ fitting. (d) FN plot of the same cycle including linear fit (red) for the high-bias data region $V\gtrsim 0.7\,\mathrm{V}$ (vertical line). The fit results in $\alpha kR=0.02\,\mathrm{nm}$, and $A=4.8\,\mathrm{nm^2}$.}
    \label{fig2}
\end{figure*}
In the following we discuss the effect of an applied voltage on the tunneling resistance. We assume that the properties of the tunneling contribution to the total conductance can be described by a tunnel barrier with an effective area $A$ and thickness $d$. The $I(V)$ characteristics of such a barrier can be approximated by~\cite{simmons63p1, simmons63p2}
\begin{align}\label{eq:simmons}
I(V)&=G_V\cdot V\left(1+\frac{1}{3}\frac{V^2}{V_0^2}\right),
\end{align}
with $G_V=A\sqrt{2m\varphi}e^2/(4\pi^2\hbar^2\cdot d)\cdot \exp(-d\sqrt{8m\varphi}/\hbar)$ and $V_0^2=4\hbar^2\varphi/e^2md^2$, where $\varphi$ denotes the work function. Relating the expression for $G_V$ to the measured resistance at small bias can be used to estimate the effective tunneling distance $d$ by making appropriate assumptions with regard to $A$ and to the ratio $p$ of the tunneling conductance to the total conduction. For the example of Fig.~\ref{fig2}(c), we obtain, with $A=3000\,$nm$^2$ and $p=0.15$, a tunneling distance of $d\approx 500\,$pm (for details see the Supplemental Material, App.~D). The value $d=500\,$pm in turn, leads to a value of the nonlinearity parameter $1/3\cdot V_0^2 \approx 0.05\,$V$^{-2}$. However, fitting Eq.~\ref{eq:simmons} to the $I(V)$ characteristics of Fig.~\ref{fig2}(c) (solid line) yields a nonlinearity parameter of $1/(3\cdot V_0^2)\approx 100\,$V$^{-2}$, i.\,e., three orders of magnitude larger than the above estimate.

This manifest contradiction can be attributed to a shrinkage of the tunnel contact distance with increasing voltage. The electrodes could approach each other, e.\,g., by thermal expansion or by stress on the atomic scale due to the electric field. Considering thermal expansion, the following expression for the voltage dependence of the resistance can been derived:
\begin{align}\label{eq:rvsv}
R(V)&=\frac{R(V=0)}{1+p\left(\exp(\beta^\ast{}V)-1\right)}
\end{align}
where $\beta^{\ast}=D\alpha_{Au}\gamma \sqrt{8m\varphi}/\hbar$ (with $D$ the typical grain size and $\alpha_{Au}$ the thermal expansion coefficient of Au) and $R(V=0)$ the resistance at zero bias. Details of the model are described in the Supplemental Material (Appendix D). A fit of Eq.~\ref{eq:rvsv} to the data reproduces the observed decrease of $R(V)$ with $\beta^{\ast}=4.00\,$V$^{-2}$ and $R(V=0)=13.7\,\mathrm{k\Omega}$, where we assumed $D\approx 30\,$nm, i.e., on the order of the initial film thickness and $\alpha_{Au}=14.2\cdot 10^{-6}\,\mathrm{K^{-1}}$~[\onlinecite{crc}], resulting in $d=450\,$pm. An analogous expression can be derived for the distance change due to electric-field mediated strain, since both mechanism to first order are linear in the voltage it is not possible to separate the two contributions. In summary, we have obtained a description of the metallic and tunneling contributions to $R(V)$ and the $I(V)$ characteristics (Fig.~\ref{fig2}(c) dashed line) by considering a change of the tunneling distance due to thermal expansion due to Joule heating and/or electrostatic strain both arising from the applied voltage. For simplicity, we have assumed in our calculation a single large tunneling contact with parameters $A$ and $d$. Of course, in reality several tunneling contacts might be present.

A similar EM thinning process has been performed and analyzed recently for metallic Permalloy (Ni$_{80}$Fe$_{20}$) films and no evidence for tunneling up to several $10\,\mathrm{k\Omega}$ have been found\cite{bieren2013}. The apparent discrepancy with the current work could be due to the larger specific resistance (by about one order of magnitude) of Permalloy, the different shape of the contact, and/or the different micro-structure compared to clean metals used here. Finally we should note that this model does not consider Coulomb-blockade effects on $R(V)$.
 
With increasing bias, the voltage dependence of the tunneling current deviates from the behavior described before, as seen in Fig.~\ref{fig2}(c) for $V\ge 0.6\,\mathrm{V}$, and can be explained by Fowler-Nordheim (FN) tunneling~\cite{fowler28, gomer61, lucier05, mueller09}:
\begin{eqnarray}
 I_{FN}= A\cdot \dfrac{e}{2\pi h}\dfrac{(\mu /\varphi)^{1/2}}{(\mu + \varphi)}\dfrac{(eV)^2}{(\alpha kR)^2} \cdot \exp \left(-\dfrac{4}{3}\cdot \frac{\sqrt{2m}}{\hbar}\cdot \varphi^{3/2}\frac{\alpha k R}{eV}\right).\label{eq3}
\end{eqnarray}
Here $A$ is the total field-emitting area and $\mu=5.53\,\mathrm{eV}$ is the Fermi level of gold~\cite{ashcroft76}. $E=V/\alpha k R$ is the electric field for an applied bias $V$ and a spherical shaped emitting area with a radius $R$. The electric field includes $\alpha$ as a correction factor to account for image charges and a field reduction factor $k$~(Ref.~\onlinecite{gomer61, lucier05}). $\alpha$ and $k$ are dimensionless. To compare our data to the Fowler-Nordheim description, we plot $\ln(I/V^2)$ versus $1/V$ in Fig.~\ref{fig2}(d). By fitting a straight line to the high-bias region $V \gtrsim 0.7\,\mathrm{V}$ ($1/V \lesssim 1.5 \cdot \,\mathrm{V^{-1}}$) we can determine $\alpha kR$ from the slope and $A$ from the interception with the axis of ordinate. The fits are limited to the small voltage range where $\ln(I/V^2)$ versus $1/V$ has a negative slope (see Eq.~(\ref{eq3})). Nevertheless the $A$ values fall in a reasonable range considering our sample geometry. When field emission sets in, material can erode from the emitter leading to melting or sublimation of material at the anode due to the bombardment by accelerated electrons. This can lead to an uncontrolled increase of the vacuum barrier, which is unfavorable for the control of electromigration. We found one example that shows two different slopes of $\ln(I/V^2)$ versus $1/V$, which might signal such an instability. We investigated its evolution by analyzing neighboring cycles as shown in the Supplemental Material (Appendix~F, Fig.~S6)~\cite{supplemental}. From these results, we ascribe sudden uncontrolled changes of $R(V)$ mostly to a change of the contact due to field emission. Hence, we tried to avoid the field-emission regime, although it was shown that nanogaps can also be realized by field-emission-induced EM~\cite{kume2010,takiya2012}. Additionally, in field-induced electron transmission between macroscopic islands, Gundlach oscillations may occur if the bias voltage exceeds the work function of the sample~\cite{gundlach66,kolesnychenko99,huisman11}. Although oscillatory behavior has been observed in some cycles (see for example Fig.~S3(a) and (b) for $V\ge 1.2\,$V and $V\ge 0.5\,$V, respectively)~\cite{supplemental}, it is not possible to attribute these features unambiguously to Gundlach oscillations.

Air components such as N$_2$, O$_2$ and water are known to change the work function at the surface of metals~\cite{eberhagen60p1}. An increase in the work function at the metal surface reduces the probability for electron tunneling, making the contribution of tunneling to the total current smaller. The work function of Au is $4.73$ to $5.26\,$eV~[\onlinecite{eberhagen60p1,landolt11p1}] in vacuum and is increased by $0.17\,$eV for H$_2$~[\onlinecite{eberhagen60p1}] and by $0.9-1.58\,$eV for O$_2$~[\onlinecite{eberhagen60p1}] absorption. It has been reported that O$_2$ and H$_2$ chemisorption has an influence on the crystallinity of Pt in the EM process, resulting in nanogaps between single-crystalline electrode tips~\cite{suga2012}. 

Also humid air is known to have an influence on EM since water is a dielectric. Moisture in copper dual-damascene interconnects of electronic devices, for example, leads to an enhanced failure probability of the device due to EM~\cite{Cheng10}. To test if negative slopes in $R(V)$, i.e., tunneling occur more frequently under UHV compared to ambient conditions, we have conducted experiments under gas dosage. Three samples were electromigrated in an oxygen atmosphere of $500\,$Pa. The samples were thinned up to $450\,\mathrm{\Omega}$ before the resistance jumped to values larger than $60\,$k$\mathrm{\Omega}$. One of these samples showed no negative $R(V)$ slopes. The other two showed only one electromigration cycle with a negative slope that was no longer observed in the further electromigration cycles. Such experiments indicate that oxygen partly inhibits the development of negative slopes. However, the contacts broke unusually early in the thinning process, therefore we conclude that oxygen also impedes controlled thinning of the contacts. The possible effect of gas dosage in samples that had already revealed negative $R(V)$ slopes was also investigated. For Ar, N$_2$ and O$_2$ with pressures between $10^{-6}$ and $10^{-2}\,$Pa no significant effect, i.e., no difference in $R(V)$ compared to the UHV EM process, was observed. H$_2$ showed no effect up to $10^{-4}\,$Pa. In some samples an effect was noticeable for H$_2$ pressure above $10^{-3}\,$Pa. However, the change occurred after $1-10\,$min and no preferred tendency to either increasing or decreasing $R(V)$ slopes was detected. Thus, the presence of H$_2$ cannot explain the systematic development of negative slopes in the $R(V)$ data.

In summary, we find that in UHV experiments the nanocontacts become sufficiently small that negative slopes in $R(V)$ appear. After comparing our data to phenomenological models, we tentatively attribute this behavior with tunneling in parallel to the ballistic transport through the nanocontact, rather than to thermal activation. Not only the negative slope in $R(V)$ but also the evidence of the transition from direct to Fowler-Nordheim tunneling supports the idea of tunneling contacts in our samples. The appearance of negative slopes in $R(V)$ is partly inhibited by oxygen dosage, but the presence of O$_2$ also leads apparently to a greater probability of breakage in the process. 
Most importantly, the possible observation of tunneling, i.e., the formation of tunneling contacts, indicates that very small gaps in the nanobridge can be formed in vacuum on a routine basis, which supports that this technique is promising to provide contacts for molecular electronics.

We thank C.\  S\"urgers for stimulating discussions and help with the experiment, and R.\ Montbrun for the preparation of the \SiN masks. Financial support of the Baden-W\"urttemberg Stiftung in the framework of its excellence program for postdoctoral fellows and from the European Research Council through the Starting Grant NANOCONTACTS (No. 239838) is gratefully acknowledged.

\end{document}